\documentclass[twocolumn,english,aps,prl,floatfix,amssymb,superscriptaddress]{revtex4-1}
\usepackage[T1]{fontenc}
\usepackage[latin9]{inputenc}
\usepackage{amsmath}
\usepackage{amssymb}
\usepackage{graphicx}

\makeatletter
\@ifundefined{textcolor}{}
{%
 \definecolor{BLACK}{gray}{0}
 \definecolor{WHITE}{gray}{1}
 \definecolor{RED}{rgb}{1,0,0}
 \definecolor{GREEN}{rgb}{0,1,0}
 \definecolor{BLUE}{rgb}{0,0,1}
 \definecolor{CYAN}{cmyk}{1,0,0,0}
 \definecolor{MAGENTA}{cmyk}{0,1,0,0}
 \definecolor{YELLOW}{cmyk}{0,0,1,0}
}

\usepackage{babel}

\usepackage{babel}

\usepackage{babel}

\usepackage{babel}

\@ifundefined{showcaptionsetup}{}{%
 \PassOptionsToPackage{caption=false}{subfig}}
\usepackage{subfig}
\makeatother

\usepackage{babel}
\begin{document}

\title{Thermodynamic properties of a quantum hall anti-dot interferometer}

\author{Sarah Levy Schreier\footnote{corresponding authors, email: sarah@sightdx.com , adiel.stern@weizmann.ac.il}}

\affiliation{Department of Condensed Matter Physics, Weizmann Institute of Science,
Rehovot, Israel 76100}

\author{Ady Stern$^{*}$}

\affiliation{Department of Condensed Matter Physics, Weizmann Institute of Science,
Rehovot, Israel 76100}

\author{Bernd Rosenow}

\affiliation{Institute for Theoretical Physics, University of Leipzig, Vor dem
Hospitaltore 1, D-04103 Leipzig, Germany}

\author{Bertrand. I. Halperin}

\affiliation{Department of Physics, Harvard University, Cambridge, Massachusetts
02138, USA}

\date{\today}
\begin{abstract}
We study quantum Hall interferometers in which the interference loop
encircles a quantum anti-dot. We base our study on thermodynamic considerations, which we believe reflect the essential aspects of interference transport phenomena. We find that similar to the more conventional
Fabry-Perot quantum Hall interferometers, in which the interference
loop forms a quantum dot, the anti-dot interferometer is affected
by the electro-static Coulomb interaction between the edge modes defining
the loop. We show that in the Aharonov-Bohm regime, in which effects
of fractional statistics should be visible, is easier to access in
interferometers based on anti-dots than in those based on dots. We
discuss the relevance of our results to recent measurements on anti-dots
interferometers.
\end{abstract}
\maketitle
\captionsetup[figure]{margin=0pt,font=normalsize,labelfont=bf, justification=raggedright}
\section{introduction}
Early after the discovery of the fractional quantum Hall states it
was realized that the charged excitations that characterize these
states carry fractional charge and satisfy fractional statistics,
abelian or non-abelian. Abelian fractional statistics is manifested
in the phase accumulated by one quasi-particle going around another \cite{LeMy77,Laughlin83,Halperin84,ArScWi84}.
The natural arena for an experimental observation of such a phase
is that of interferometry. Consequently, large experimental and theoretical
effort has been devoted to the study of quantum Hall interferometers
of various types, such as the Fabry-Perot \cite{Chamon+97}   and the Mach-Zehnder \cite{Chung+03} interferometers.
In these interferometers, current is introduced from a source and
is distributed between two drains. The relative distribution between
the drains involves interference between trajectories that go around
an interference loop.

In an idealized model, when an integer quantum Hall state is examined,
a variation of the magnetic field continuously varies the flux in
the interference loop thus leading to Aharonov-Bohm oscillations. In fractional quantum Hall states another factor becomes important:
the variation of the flux affects the number of localized quasi-particles
within the loop. Since at low temperature this number is quantized
to an integer, and since the mutual statistics of the interfering quasi-particle with the localized one is fractional, for abelian quantum Hall states the introduction of
a localized quasi-particle into the bulk leads to a phase jump. As
the temperature is raised the phase jump is expected to gradually
smear.

A major deviation from the idealized model occurs in a Fabry-Perot
interferometer due to the capacitive coupling between the interferometer's
edge and bulk. As a result of this coupling, the variation of the
magnetic field varies also the area of the interferometer, thus complicating
the dependence of the Aharonov-Bohm phase on the field \cite{RoHa07}. Furthermore,
the introduction of a quasi-particle into the loop leads also to a
sharp change in the area, affecting the discontinuous jump in the
phase. The bulk-edge capacitive coupling was analyzed in details in
\cite{halperin2011theory}, where a distinction has been made between
the Aharonov-Bohm (AB) case, where the capacitive coupling is weak
and the idealized model is a good approximation, and the Coulomb-dominated
(CD) case, where the capacitive coupling is strong and the idealized
model does not hold. This distinction holds both in the limit in which
the interferometer is fairly open, and only two trajectories interfere,
and in the limit of a closed interferometer (a quantum dot), in which
multiple reflections are important, and the sinusoidal interference
patterns are replaced by resonances.

In this work we examine interference in systems based on an anti-dot
(AD) embedded in a bulk that is in a quantum Hall state  \cite{GoSu95,Goldman+08,ihnatsenka2009electron,LeeSim10,Ilan+11,kou2012coulomb,Paradiso+12,Martins+13,Martins+13a,Pascher+14, Kataoka +00, Geller+97, Geller+00, Braggio+06, Merlo+07} . A typical
set-up is depicted in Fig. (\ref{schematic anti-dot}) \footnote{Ref. \cite{halperin2011theory}'s terminology relates to ours as follows:
$\nu_{in}=\nu_{1}$ and $\nu_{out}=\nu_{2}$.}. The anti-dot is a region where the density is fully depleted. As
a consequence, it is encircled by edge modes of finite size. When
the antidot is weakly coupled to the two ends of a quantum point contact,
it induces back-scattering of the current in a quantum Hall device,
provided that it is not blockaded by Coulomb charging energy. The
AD may be tuned between transmission resonance peaks and dips by
means of a magnetic field or gate voltage. We explore the dependence
of the Coulomb blockade peaks on the magnetic field and on a gate
voltage applied to the antidot, and analyze the information contained
in this dependence on the mutual statistics of quasi-particles. We
find that, when compared to a Fabri-Perot interferometer,  the antidot geometry is potentially much easier to drive
into the AB regime in which this information is accessible. Although the analysis we carry out is of thermodynamic quantities, namely the equilibrium charge on the anti-dot, we expect its essential features to be reflected also in transport, due to the coupling between transport and thermodynamics in the Coulomb blockade regime.

\section{The Model}

\begin{figure}
\subfloat[]{\protect\includegraphics
[scale=0.15]
{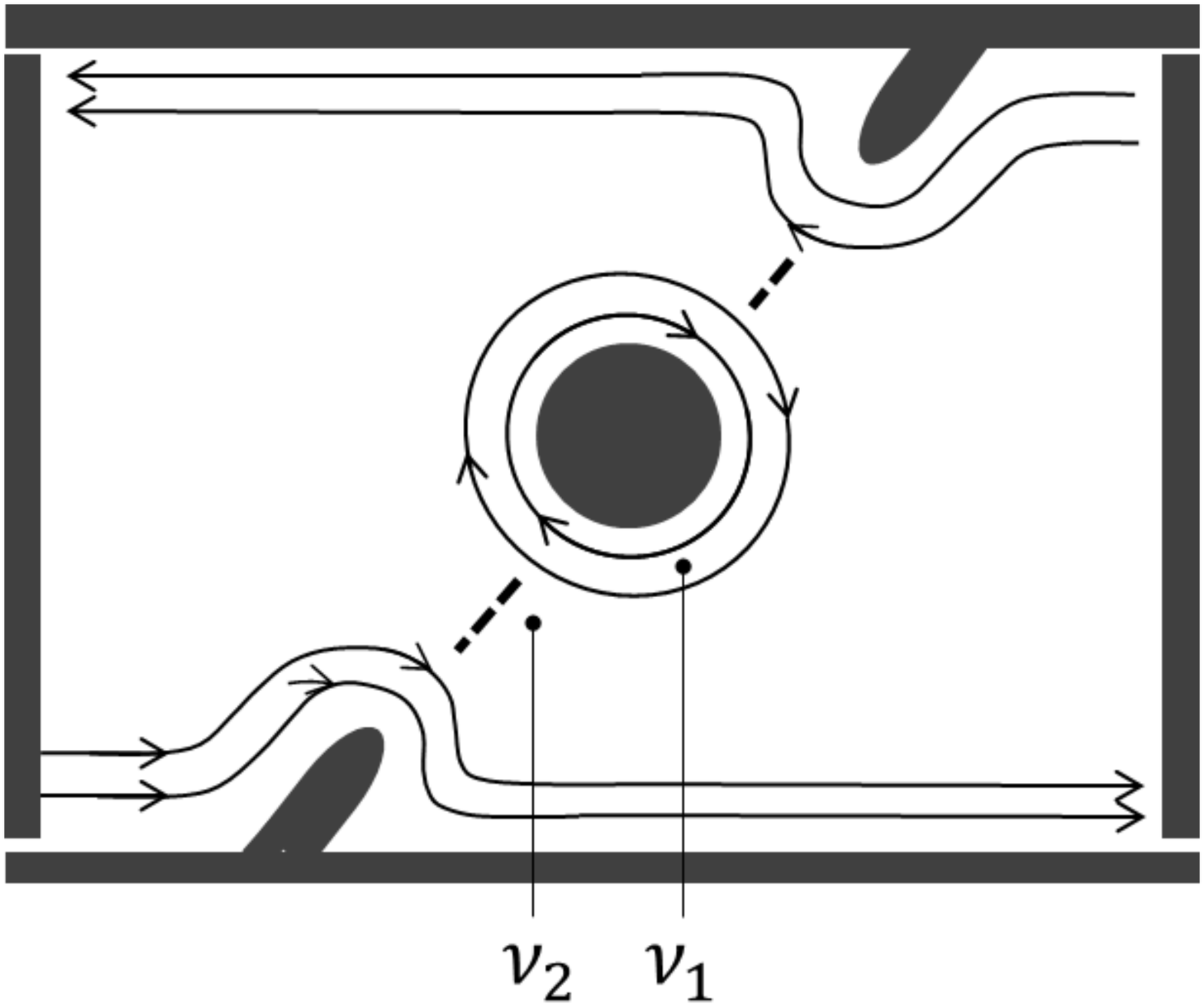}

}~~\subfloat[]{\protect\includegraphics
[scale=0.15]
{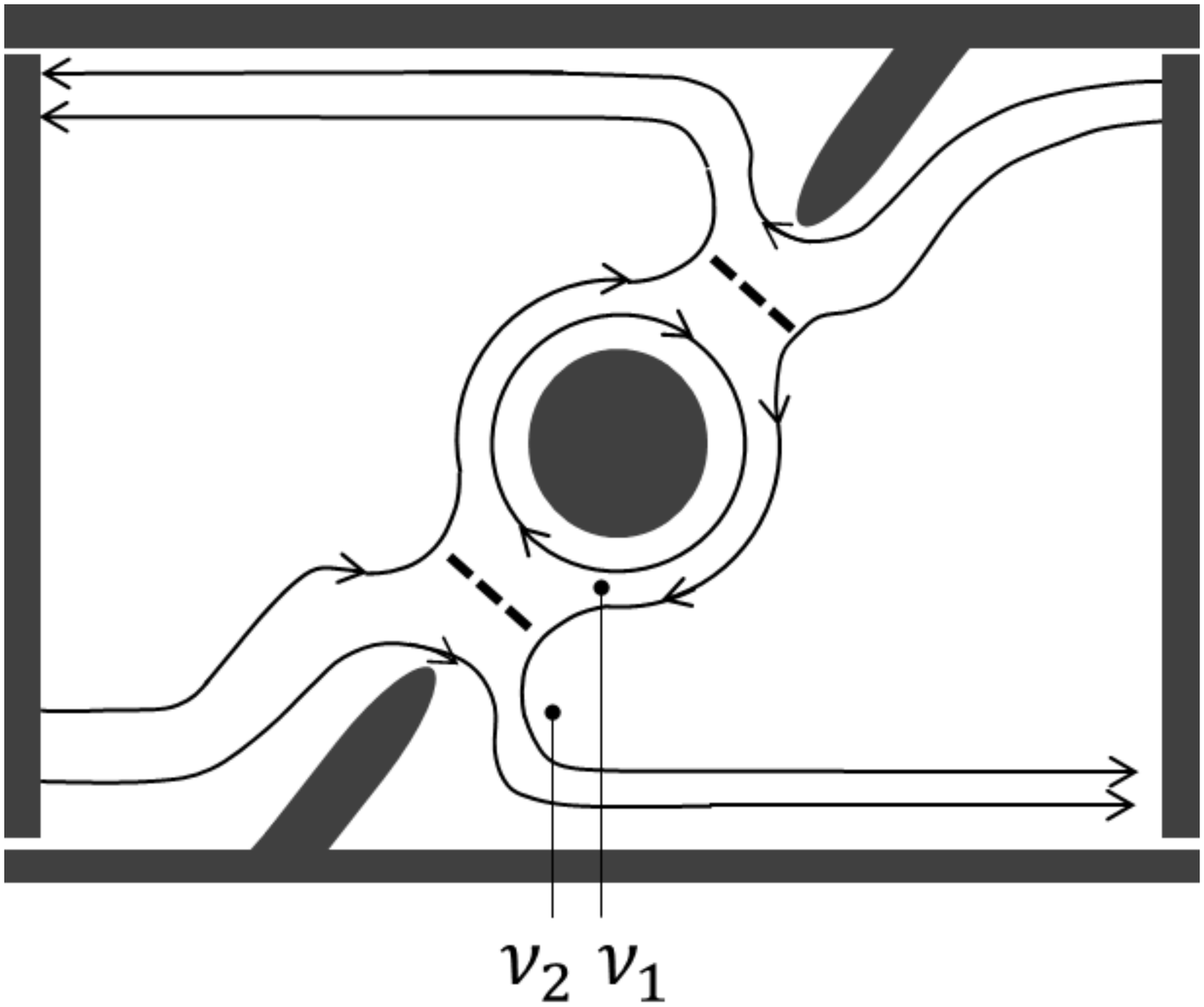}

}

\subfloat[]{\protect\includegraphics
[scale=0.15]
{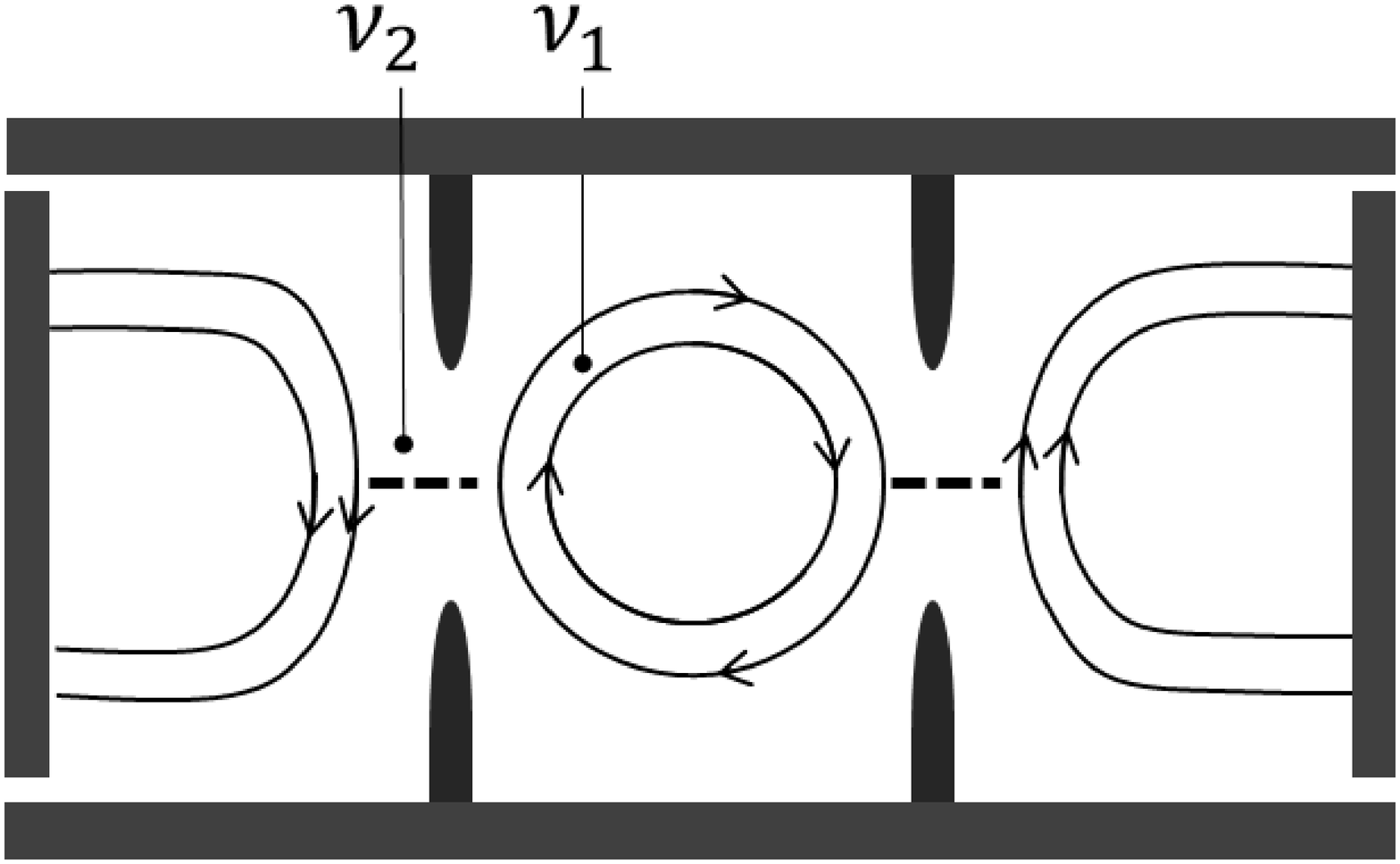}

}~~\subfloat[]{\protect\includegraphics
[scale=0.15]
{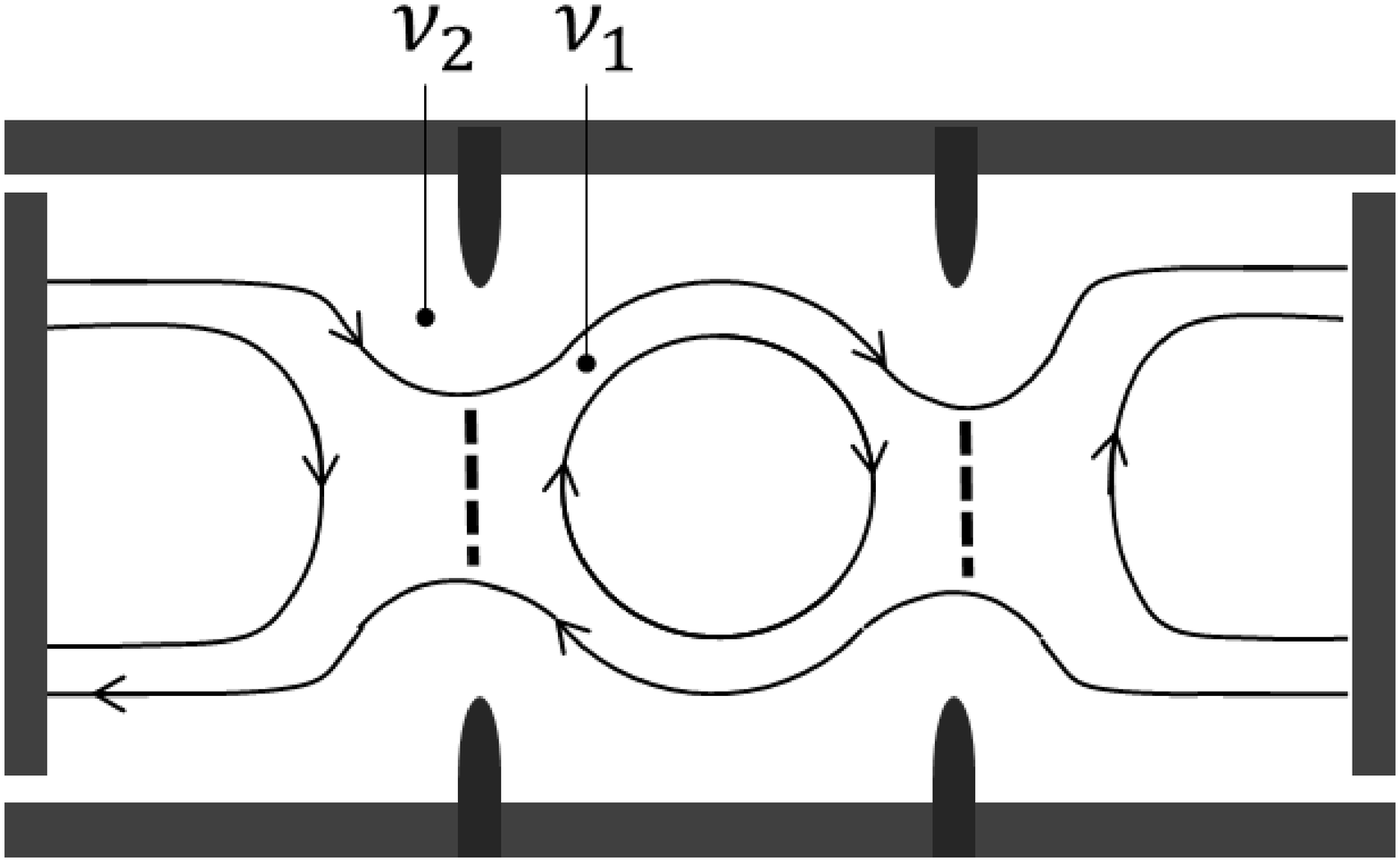}

}

{\large{}\protect\caption{\label{schematic anti-dot}{\footnotesize (a)\&(b) Schematic layouts of the edge states through AD based devices. The partially transmitted edge channel separates two quantized Hall states corresponding to rational filling factors $\nu_{1}$ and $\nu_{2}$, with $\nu_{1}$ being closer to the island. (a) Anti dot in the weak back scattering limit, where the constrictions filling factor is given by $\nu_{2}$. (b) Anti dot in the strong back-scattering limit, where the constrictions filling factor is given by $\nu_{1}$. (c)\&(d) Schematic layouts of a FPI with a single partially transmitted edge channel penetrating the two constrictions. The partially transmitted edge channel separates two quantized Hall states corresponding to rational filling factors $\nu_{1}>\nu_{2}$, with $\nu_{2}$ being closer to the sample edge. (c) FPI in the strong back-scattering limit, where the constrictions filling factor is given by $\nu_{2}$. (d) FPI in the weak back-scattering limit, where the constrictions filling factor is given by $\nu_{1}$. The most probable tunneling paths across the constrictions, in each configuration, are indicated by dashed lines. The anti dot configurations (a) and (b) represent the counter parts of the FPI configurations (c) and (d), respectively.}}
}{\large \par}
\end{figure}

In the case of weak coupling of the AD to the Quantum point contacts (Fig.\ref{schematic anti-dot}(a)),
the charge on the anti-dot is approximately quantized in units of
the charged excitations of the quantum Hall state that surrounds the
antidot. Resonant backscattering through the edge modes encircling the AD takes
place only when there is a degeneracy of the ground state energy of
the antidot for two values of the charge. To specify that energy we
define a model for the antidot. We focus on the case of two edge modes
surrounding the antidot, denoted by 1 and 2 (inner and outer edges,
respectively). The filling factor at the constriction is denoted by
$\nu_{2}$. The antidot, being depleted, is at filling
factor zero. The filling factor between the two edge modes is denoted
by $\nu_{1}$. The outer edge channel separates two quantized Hall
states corresponding to rational filling factors $\nu_{1}$ and $\nu_{2}$.
The states $\nu_{1}$ and $\nu_{2}$ are either integer quantized
Hall (IQH) states, or fractional quantum Hall (FQH) states described
in the Composite Fermions picture. The corresponding quasi-particle
excitations are given by $e_{1}$ and $e_{2}$. In the T=0 limit,
the total charge in the AD island is quantized in units of $e_{2}$.


We first assume that there is no tunneling of charge between the two
edge modes around the antidot, so that charge is quantized in each
of the modes. We denote by $N$ the total number of quasi-particles
on the two closed edge channels, and by $N_{1}$ the number of quasi-particles
on the inner edge channel. Since $N$ and $N_{1}$ depend only on
what happens on the two encircling edge channels, we define an energy
functional, $E(N,N_{1})$, to be the total energy of the system when
$N$ and $N_{1}$ are specified, minimized with respect to all other
dynamical variables. In the absence of tunneling between the two edge modes
we can approximate the energy of the modes by a quadratic form in
terms of $N,N_{1}$:
\begin{equation}
E(N,N_{1})=\frac{K_{1}}{2}\left(\delta n_{1}\right)^{2}+\frac{K_{2}}{2}\left(\delta n_{2}\right)^{2}+K_{12}\delta n_{1}\delta n_{2}\label{Energy function-1}
\end{equation}
where $\delta n_{i}$ ($i=1,2$) is the deviation of the charge on
the $i$th interfering AD trapped edge channel in units of the electron
charge $e$, and given by:
\begin{eqnarray}
\delta n_{1} & = & N_{1}e_{1}+\phi_{1}\Delta\nu_{1}-\bar{q}_{1}\label{eq:delta n1}
\end{eqnarray}
\begin{equation}
\delta n_{2}=(Ne_{2}-N_{1}e_{1})+\phi_{2}\Delta\nu_{2}-\bar{q}_{2}\label{eq: delta_n2}
\end{equation}
The flux through the closed interfering edges is given by $\phi_{i}=\frac{BA_{i}}{\phi_{0}}$,
with $A_{i}$ being the area enclosed by the $i$th interfering edge.
The terms $\phi_i\Delta\nu_i$ (with $i=1,2$) are related to the quantized Hall conductance
of the different incompressible regions.
When the flux changes, the resulting
deviation of charge on the edge is  given by $\phi_{i}\Delta\nu$,
with $\Delta\nu_{1}=\nu_{1}-0$ and $\Delta\nu_{2}=\nu_{2}-\nu_{1}.$
The quantities $\bar{q}_{i}$ describe the effect of the positive background on the equilibrium charge density distribution.
The variation of $\bar{q}_{i}$ and the area $A_{i}$
as $V_{g}$ is being varied depends on the coupling of the gate to
the interferometer, and is characterized (following \cite{halperin2011theory})
by two parameters:
\begin{equation}
\begin{split}\beta_{i}=\left(\frac{B}{\phi_{0}}\right)\frac{dA_{i}}{dV_{g}}, & \,\,\gamma_{i}=\frac{d\bar{q}_{i}}{dV_{g}}\end{split}
\label{eq:beta and gamma}
\end{equation}

We consider the case of an ideal side gate, where the effect of the gate voltage $\delta V_{g}$
is to alter the location of the edge, without changing its density
profile \cite{McCLure12}. In the limit of a sharp density profile around the AD , $\gamma_{i}$
is proportional to the distance between two adjacent edges ($\delta R_{i}$)
while $\beta_{i}\bar{\nu_{i}}$ (where $\bar{\nu}_{i}\equiv{\frac{n\phi_{0}}{B}}$, $n$ is the bulk electron density
and $\bar{\nu_{i}}\geq\nu_{i}$ in the Coulomb blockade limit) is
proportional to the radius $R_{i}$. Hence, for an AD configuration
in the Coulomb blockade limit we assume:
\begin{equation}
\gamma_{i}\ll\beta_{i}\nu_{i}\label{eq:gamma << beta}
\end{equation}
The coefficients $K_{1}$ and $K_{2}$ are related to the capacitance
of the respective edges to ground (denoted by $C_{1}$ and $C_{2}$),
as described in details in \cite{RoHa07,halperin2011theory}. For the state
to be stable, they must satisfy $K_{1}K_{2}>K_{12}^{2}$. For a steep
density profile at the edge, we estimate: $A_{2}=A_{1}+\delta A$,
leading to $C_{1}=C_{2}+\delta C$, where $\frac{\delta C}{C_{1}}\ll1$,
$\frac{\delta A}{A_{2}}\ll1$, so that in general $K_{1}\approx K_{2}.$
The mutual coupling term $K_{12}$ describes the coupling between
the two edges. We consider the case where $K_{12}>0$ such that there is a repulsive interaction between excess charges on the two edges.   For brevity, we present here a calculation done under the assumption $K_{1}=K_{2}$,
and generalize to $K_{1}\approx K_{2}$ in the Figures. When $K_{1}=K_{2}$
the macroscopic energy function is:

\begin{equation}
E=\frac{K_{A}}{2}(\delta n_{1}+\delta n_{2})^{2}+\frac{K_{B}}{2}(\delta n_{1}-\delta n_{2})^{2}\label{eq:macroscopic energy function}
\end{equation}
where $K_{1}\approx K_{2}=(K_{A}+K_{B})$, $K_{12}=(K_{A}-K_{B})$.
Assuming the mutual capacitance to be primarily electrostatic, the repulsive interaction between electrons leads to $K_{A}\geq K_{B}$. In this form the term $(\delta n_{1}+\delta n_{2})$
represents the total charge in the AD island, and the term involving $(\delta n_{1}-\delta n_{2})$
represents the energy cost associated with charge imbalance between
the two modes encircling the AD. Since in the quantum Hall effect
charge translates into distance, this term represents the dependence
of the energy on the distance between the edge modes. A previous theoretical
model \cite{ihnatsenka2009electron} analyzed interference in ADs
in a similar way, but dealt only with the limit where $K_{B}=0$,
and neglected the significant effect of nonzero $K_{B}$ presented
here.

\section{Magnetic field and Gate voltage periodicities}
We now address the magnetic field dependence of the backscattered
current as the capacitive coupling varies between $K_{B}=0$, the
extreme \emph{Coulomb dominated} (CD)\emph{ }limit, and $K_{A}=K_{B}$,
the extreme \emph{Aharonov Bohm }(AB) limit. This definition of limits
is inherited from the model \cite{halperin2011theory}. At zero temperature,
resonant transmission occurs when there is a vanishing energy cost
for varying $N$ by one, i.e for adding one electron to one of the
two closed edges. In the extreme CD regime, the degeneracy points
are separated on the B axis by spacings: $\Delta B=\frac{\phi_{0}}{A}\frac{e_{2}}{\nu_{2}}$
(Here and below we assume that the difference between the areas $A_1,A_2$ enclosed by the two edges is much smaller than their average, approximate them to be equal, and denote the approximate area by $A$). In the extreme
AB limit the degeneracy points are two periodic structures superimposed
on one another and the values of $\bar{q}_{1}$ and $\bar{q}_{2}$
determine their relative separation.

In the
high temperature limit (i.e $T\gg K_{1},K_{2},K_{12}$) the magnetic
field and voltage dependence of the thermal average of $N$, denoted
by $\langle N\rangle$, includes oscillatory components whose magnitude
is exponentially suppressed with temperature. These oscillations may
be directly probed by a thermodynamical measurement, e.g., that of
the dot's compressibility. Alternatively, they may be probed in transport,
since current flows through the dot when $\langle N\rangle$ is close
to half integer. Notably, transport through the antidot is affected by other factors beyond $\langle N\rangle$, such as the temperature dependence of the edge propagators, and those factors are left out here.
However, we expect
our main conclusion, regarding the separation between the AB and CD
regimes, to be qualitatively applicable to transport measurements
as well.

The thermal average satisfies
\begin{equation}
\langle N\rangle=\frac{1}{Z}\sum_{N_{1}}\sum_{N}e^{-E/T}N\label{eq:<N> definition}
\end{equation}
Using the Poisson Summation Formula (details of the method are given in  \cite{halperin2011theory})
we obtain:
\begin{equation}
Z=\sum_{N_{1}}\sum_{N}e^{-E(N,N_{1})/T}=Z_{0,0}+\sum_{\left\{ g_{1},g_{2}\neq0\right\} }Z_{g_{1},g_{2},}
\end{equation} where:
\begin{equation}
Z_{g_{1},g_{2}}=\int_{-\infty}^{\infty} dN_{1} \int_{-\infty}^{\infty} dN~~e^{-\frac{1}{T}E(N,N_{1})}e^{-2\pi ig_{1}N_{1}}e^{-2\pi i g_{2}N}
\end{equation}

\begin{figure}
\subfloat[]{\protect\includegraphics
[scale=0.3]
{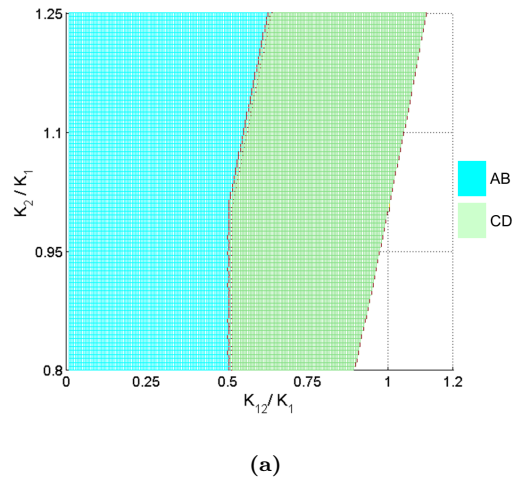}

}

\subfloat[]{\protect\includegraphics
[scale=0.3]
{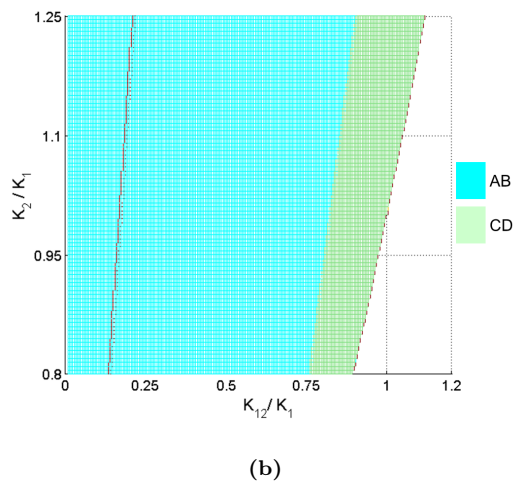}

}

\protect\caption{\label{fig:Kab2Ka =000026 Kb2Ka}{\footnotesize The separation into domains in the high temperature limit, as a function of $\frac{K_{12}}{K_{1}}$ and $\frac{K_{2}}{K_{1}}$ in the parametrization given by \ref{Energy function-1}. The left side of the plot (small $K_{12}$) corresponds to the extreme AB regime, and the right side (large $K_{12}$) to the extreme CD regime. The white area is not accessible due to the stability criterion $K_{1}K_{2}\geq K_{12}^{2}$. The colors represent the separation into domains calculated for anti-dots. The dotted line represents the border between the two regimes calculated for FPIs. (a) IQH state. AD: $\nu_{1}=1$, $\nu_{2}=2$; FPI: $\nu_{1}=2$, $\nu_{1}=1$. The separation line for the FPI is identical to that of the AD. (b) FQH state. AD: $\nu_{1}=\frac{1}{3}$, $\nu_{2}=\frac{2}{5}$; FPI: $\nu_{1}=\frac{2}{5}$, $\nu_{2}=\frac{1}{3}$. The AB regime is much larger for antidots.}}
\end{figure}

The formulas simplify at high temperatures, where the partition function
(in the denominator of Eq. (\ref{eq:<N> definition})) becomes independent of $\bar{\phi}$ (where $\bar{\phi}$
is the flux through the two closed interfering edges, $\phi_{1}\approx\phi_{2}\equiv\bar{\phi}$
in the limit discussed)
and we may focus
on the numerator of (\ref{eq:<N> definition}) and write:
\begin{equation}
\langle N\rangle=\bar{N}+\sum_{g_{1}}\sum_{g_{2}}D(g_{1},g_{2})e^{-2\pi i\left(g_{1}\bar{N}_{1}+g_{2}\bar{N}\right)}=\bar{N}+\langle\delta N\rangle\label{eq:<N> stage 1}
\end{equation}
Here, $\bar{N_{1}}=-\frac{\nu_{1}}{e_{1}}\bar{\phi}+\frac{\bar{q}_{1}}{e_{1}}$
and $\bar{N}=-\frac{\nu_{2}}{e_{2}}\bar{\phi}+\frac{(\bar{q}_{1}+\bar{q}_{2})}{e_{2}}$
.
 Furthermore,
\begin{equation}
D(g_{1},g_{2})=\frac{2\pi iT\left(4g_{2}e_{1}^{2}+2g_{1}e_{1}e_{2}\right)}{4K_{A}e_{1}^{2}e_{2}^{2}}e^{-2\pi^{2}T\frac{1}{E_{w}(g_{1,}g_{2})}}
\end{equation}
and
\begin{equation}
\frac{1}{E_{w}(g_{1},g_{2})}=\frac{1}{4e_{1}^{2}e_{2}^{2}}\left[\frac{\left(2e_{1}g_{2}+e_{2}g_{1}\right)^{2}}{K_{A}}+\frac{e_{2}^{2}g_{1}^{2}}{K_{B}}\right]\label{eq:1/E_w}
\end{equation}
In the high temperature limit $D(g_{1},g_{2})$ is dominated by the
$g_{1},g_{2}$ values for which $1/E_{w}(g_{1},g_{2})$ is smallest.
In the AD we always have $e_{1}\geq e_{2}$, hence the AB magnetic
field periodicity originates from $g_{1}=\pm1,\,\,g_{2}=0$ and dominates
when $\frac{K_{B}}{K_{A}}\rightarrow1$. The CD magnetic field periodicity
originates from $g_{1}=0,\,\,g_{2}=\pm1$ and dominates when $\frac{K_{B}}{K_{A}}\rightarrow0$.
The transition between the two occurs at:
\begin{equation}
\frac{K_{B}}{K_{A}}=\left(4\frac{e_{1}^{2}}{e_{2}^{2}}-1\right)^{-1}\label{eq:KB/KA condition}
\end{equation}
as demonstrated in figure (\ref{fig:Kab2Ka =000026 Kb2Ka}) (a)
and (b) for filling factors $\nu_{2}=2$ and $\nu_{2}=2/5$, respectively.
The corresponding magnetic field periodicities are given by:
\begin{equation}
\Delta B=\begin{cases}
\begin{array}{cc}
\frac{e_{2}}{\nu_{2}}\frac{\phi_{0}}{A} \,&\,\, {\rm for\,\, CD\,\, domain}\\
\\
\frac{\phi_{0}}{A}\, &\,\, {\rm for\,\,AB\,\, domain}
\end{array}
\end{cases}
\end{equation}
where we substitute $\nu_{1}/e_{1}=1$, which is always true as long
as we discuss fractional quantum Hall states that may be mapped onto
integer quantum Hall states of Composite Fermions. At fractional filling
factor $\nu_{2}=\frac{2}{5}$ , as is evident in figure (\ref{fig:Kab2Ka =000026 Kb2Ka}), the Coulomb dominated\emph{ }regime
($g_{1}=0,$$g_{2}=\pm1$) is much smaller than at integer filling
factor $\nu_{2}=2$, and the system is mainly in the\emph{ }Aharonov
Bohm domain ($g_{1}=\pm1,\,g_{2}=0$).

Equation (\ref{eq:1/E_w}) demonstrates the crucial role that the
ratio $e_{1}/e_{2}$ plays here. For antidots this ratio is always
$\geq1$, leading to a smaller CD domain at fractional filling factors.
For Fabry-Perot interferometers the corresponding ratio $\frac{e_{1}}{e_{2}}\leq1$,
hence at fractional filling factors the system is predominantly in
the CD domain. This result suggests that it might be easier to explore
an AB behavior in AD interferometers than in FP interferometers.

Figure \ref{fig:Kab2Ka =000026 Kb2Ka} maps the AB regime, the CD
regime and the transition line as a function of the ratios $K_{12}/K_{1}$
and $K_{2}/K_{1}$. The figure also shows the transition line between
the two regimes for FPIs. The position of the line shows that even when $K_2,K_1$ are not equal to one another, the AB region for the anti-dot case is significantly larger than for the FPI case.
We note, however, that when $K_{2}/K_{1}$ is very different from
one and when the energy functional (\ref{eq:macroscopic energy function})
is close to an instability new regimes emerge, with different dominant
values of $g_{1},g_{2}$.

We now use the same approach to extract the gate voltage periodicity.
For a given $\delta V_{g}\neq0$ and fixed magnetic field we obtain
from equation (\ref{eq:<N> stage 1}) the gate voltage periodicity
in the CD and AB limits:
\begin{equation}
\Delta V_{g}^{\,(CD)}=\frac{e_{2}}{\beta\nu_{2}},\,\,\Delta V_{g}^{\,(AB)}=\frac{e_{1}}{\beta\nu_{1}}\label{eq:DeltaVg}
\end{equation}
 (here $\beta_{i}=\frac{B}{\phi_{0}}\frac{dA_{i}}{dV_{g}}$ and we approximate $\beta_1\approx\beta_2$ and denote both by $\beta$; Furthermore, we use the limit (\ref{eq:gamma << beta})).
It is interesting to explore the ratio between the expressions given
in (\ref{eq:DeltaVg}) and the gate voltage periodicity $\Delta V_g^{(1)}=\frac{\phi_{0}}{B_{0}}\left(\frac{dA}{dV_{g}}\right )^{-1}$ that holds when the filling factor at the constrictions of the Anti Dot is one and $e_{1}=1$. 
Here $B_{0}$ is the magnetic field when the filling factor in the constriction is one, and generally $B=\frac{B_{0}}{\nu_{2}}$.
Then,
\begin{equation}
\frac{\Delta V_{g}}{\Delta V_{g}^{(1)}}=\begin{cases}
\begin{array}{cc}
e_{2} \,&\,\, {\rm for\,\, CD\,\, domain}\\
\\
\frac{B_{0}}{B}=\nu_{2} \,&\,\, {\rm for\,\, AB\,\, domain}
\end{array}
\end{cases}\label{eq:Delta Vg ratios}
\end{equation}
Again we assumed that the voltage is applied by an ideal side gate, such that $\frac{dA}{dV_{g}}$ does not depend on filling factor \cite{McCLure12}.

As Eq. (\ref{eq:Delta Vg ratios}) shows, in the CD domain the voltage-periodicity ratio is the quasiparticle charge, while in
the AB domain it is the ratio between the magnetic fields. Let us
examine this result at $\nu_{2}=2$ and $\nu_{2}=\frac{2}{5}$. In
the CD regime the ratio equals the charge of the tunneling quasi-particle, which is
$e_{2}=1$ or $\frac{1}{5}$ respectively. In contrast, in the AB domain the
ratio is twice as large and equals $2$ or $\frac{2}{5}$, respectively.

The accessibility of the Aharonov-Bohm regime in anti-dot interferometers
may be expected to bear on the visibility of effects of fractional
statistics in these interferometers \cite{halperin2011theory} (see also \cite{Levkivskyi}). In
this regime the two closed edge channels are independent, therefore
the only influence that the number $N_{1}$ of quasiparticles on the inner edge can have on the outer edge is through mutual statistics. At the zero temperature limit, in the AB domain, we seek to minimize simultaneously the expressions $(\delta n_{1})^2=(N_{1}e_{1}+\bar{\phi}\nu_{1}-\bar{q}_{1})^2$
and $(\delta n_{2})^2=\left [(Ne_{2}-N_{1}e_{1})+\bar{\phi}(\nu_{2}-\nu_{1})-\bar{q}_{2}\right ]^2$ and we identify the Coulomb blockade peaks in which two consecutive values of $N$ result in the same energy. For a fixed ${\bar q}_i$ the minimization of $(\delta n_2)^2$ implies that the flux separation $\Delta\bar{\phi}$ between two such Coulomb blockade peaks satisfies:
\begin{equation}
1=e_{1}\left(\Delta N_{1}\frac{1}{e_{2}}+\Delta\bar{\phi}\right)\label{shifts}
\end{equation}
where we used the relation $(\nu_{2}-\nu_{1}) =e_{1}e_{2}$.
 $\Delta N_{1}$ is the discrete change in the equilibrium value of $N_{1}$ between two consecutive peaks.
As long as $\Delta N_{1}=0$, we have $\Delta\bar{ \phi} = 1/e_1$. However, the minimization of $(\delta n_1)^2$ enforces  discrete changes in $N_{1}$ that
lead to a shift in the resonant transmission peaks as a function of $\bar{\phi}$.
These shifts are a consequence of the mutual statistics between the
quasi-particles. For IQH states, $\frac{e_{1}}{e_{2}}=integer$, the
shifts are unnoticeable and the magnetic field periodicity stays $\frac{\phi_{0}}{A}$,
as expected. Notably, even for FQH states we find that as the magnetic field is varied $N_{1}$ changes
in such a way that the magnetic field separation between peaks is $\frac{\phi_{0}}{A}$, which is the fundamental AB periodicity. This is a result of energy considerations that force the equilibrium values of the charges on both edges to change as the flux changes

The restoration of the fundamental Aharonov-Bohm periodicity $\phi_{0}/A$
is a somewhat disappointing consequence of the fractional statistics.
A more striking consequence may be seen in the limit in which the
anti-dot is strongly coupled to the edges (Fig. (1b)). Even at low temperature, most characteristics
of that limit are similar to those of the high temperature limit of
the weak coupling case. However, in the present context it is important
to note that the strong coupling to the edge makes the magnetic field
dependence of the conductance sinusoidal, and a change in $N_{1}$
leads to a phase jump, due to the mutual statistics of quasi-particles
in the inner and outer edges. An observation of these phase jumps
is an observation of fractional statistics \cite{jain1993proposed}.
Signatures of such phase jumps may have been experimentally observed
by An et al \cite{an2011braiding}. An interesting consequence of
the fact that $\frac{e_{1}}{e_{2}}\geq1$ is that in the strong coupling
limit, in between the two limiting AB and CD cases, intermediate regimes
may occur as well, with different flux periodicities.

\section{The case of $\nu=2/3$}
We conclude by examining the case of $\nu=2/3$, for which experimental
results are published \cite{kou2012coulomb}. This state is characterized
by a co-existence of two counter-propagating
edge modes. Under the assumption of sharp confinement, the spin-polarized
$\nu=2/3$ state is supposed to include an edge mode of $\nu_{1}=1$
with a charge of $e_{1}=1$, and a counter-propagating edge mode of
$\nu=-1/3$ with a charge of $e_{2}=-1/3$. At long distances tunneling of electrons between the two edges mixes them into charged and neutral modes. For a small sized anti-dot, we assume that such tunneling is negligible, in which case our analysis above applies in
a straight-forward manner.

\begin{figure}
\includegraphics
[scale=0.4, trim=0 200 0 200]
{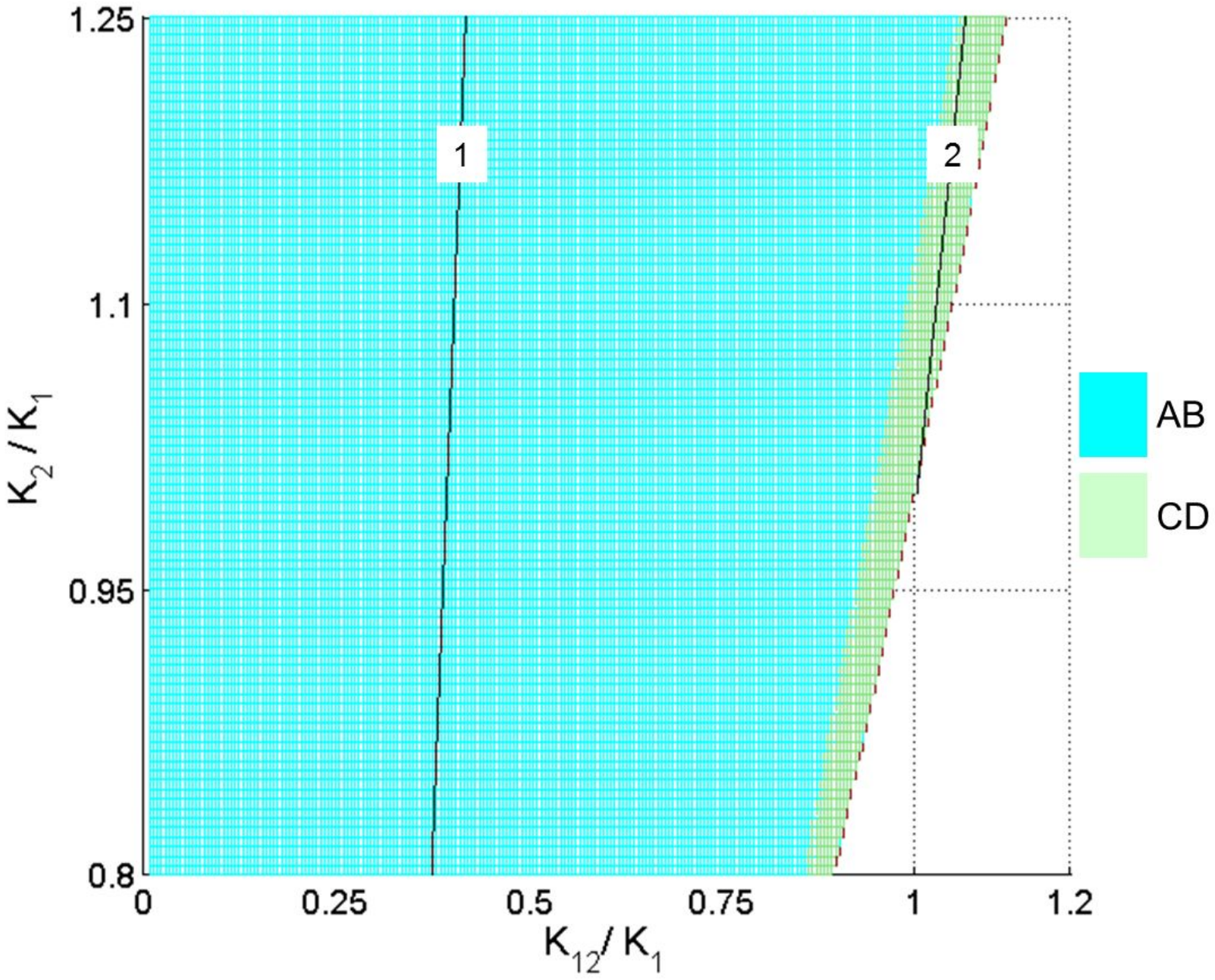}

\protect\caption{\label{fig:2-3}{\footnotesize The separation into domains in the high temperature limit, as a function of $\frac{K_{12}}{K_{1}}$ and $\frac{K_{2}}{K_{1}}$ in the parametrization given by \ref{Energy function-1}. The left side of the plot (small $K_{12}$) corresponds to the extreme AB regime, and the right side (large $K_{12}$) to the extreme CD regime. The colors represent the separation into domains calculated for anti-dots at $\nu_{2}=\frac{2}{3}$ ($\nu_{1}=1$). The two lines (denoted by 1,2) mark the domain where $\Delta=\frac{3}{2}$ (lines 1) and where $\Delta=1$ (line 2).}}
\end{figure}

Figure \ref{fig:2-3} shows that at $\nu_{2}=\frac{2}{3}$ the CD
regime is much smaller than at $\nu_{2}=2$. Even a small non-zero
value of $K_{B}$
($\sim$30 times smaller than $K_{A}$)
is enough to change the dominant periodicity from a CD periodicity
at integer filling factor, to an AB dominated one at $\nu_{2}=\frac{2}{3}$.
This result suggests a new interpretation to the experimental measurements
of magnetic field and gate voltage periodicities at $\nu_{2}=\frac{2}{3}$
and $\nu_{2}=2$ published by Kou et al. \cite{kou2012coulomb}. In
this interpretation the measurements at $\nu_{2}=2$ are in the CD
domain, while those at $\nu_{2}=\frac{2}{3}$ are in the AB domain.
Then, the result $\frac{\Delta V_{g}(\nu_{2}=\frac{2}{3})}{\Delta V_{g}(\nu_{c}=1)}=\frac{2}{3}$
reflects the ratio between magnetic fields, rather than the tunneling
charge (as obtained in equation (\ref{eq:Delta Vg ratios})). This conclusion
relies on the assumption that $K_{B}\neq0$, which is necessary to
ensure that the two counter-propagating edge channels stay separated.

When considering electron tunneling between the two edge modes we
are led to two observations. First, the quantization of the conductance
at $\nu=2/3$ to $\frac{2e^{2}}{3h}$, observed experimentally in
large samples at the lowest accessible temperatures, indicates that
tunneling between the edge modes is a relevant perturbation \cite{Kane1994}.
The relevance of this perturbation is determined by the relative strength
of the intra- and inter- edge interaction. These are parameterized
in \cite{Kane1994} by the velocity parameters $v_{1},v_{2},v_{12}$,
and the condition for relevance is expressed in terms of the parameter
$\Delta$:
\begin{equation}
  \Delta=\frac{2-\sqrt{3}C}{\sqrt{1-C^2}}
\end{equation}
where:
\begin{equation}
 C=\frac{2v_{12}}{\sqrt{3}}(v_{1}+v_{2})^{-1}
\end{equation}
Translated to our notation the velocity parameters are
expressed in terms of $K_{1},K_{2},K_{12}$ as follows:
\begin{equation}
  \begin{aligned}
        v_{1}=\frac{K_{1}}{2}    \\
        v_{2}=\frac{K_{2}}{6}   \\
        v_{12}=\frac{K_{12}}{2}
  \end{aligned}
\end{equation} The region in which inter-edge scattering is relevant (e.g., $\Delta<\frac{3}{2}$) is the region to the right of line $1$ in Fig. (\ref{fig:2-3}). This observation shrinks the AB region
of the parameter space, but still leaves it significantly larger than
the CD region.

The second observation deals with line number $2$ in Fig. (\ref{fig:2-3}).
This line is the line in which the edge velocity matrix is diagonal
in the basis of charged and neutral edges, with the two edges having
velocities of $v_{c},v_{n}$. Along this line, the effect of inter-edge
scattering may be treated exactly and shown not to shift the position
of the system in the two-parameter plane of Fig. (\ref{fig:2-3}).
However, as seen in Fig. (\ref{fig:2-3}), at least as long as $K_{1}$
and $K_{2}$ are not too different, the line $2$ lies almost entirely
within the CD regime. Thus, it seems that a picture based on the decoupling
of a charge and a neutral mode for $\nu=2/3$ is less likely to explain
the observed behavior for $\nu=2/3$. The finite size of the dot may
be the reason for why the RG flow towards the decoupled fixed point
doesn't reach the fixed point itself.

\section{Conclusion}
In summary, we analyzed in this paper quantum Hall interferometers
based on quantum anti-dots. We showed that for fractional quantum
Hall states such interferometers are more likely to exhibit the Aharonov-Bohm
regime than those based on quantum dots. From that point of view, they hold more promise for the observation of fractional statistics than interferometers based on quantum dots. We argued that our observations
may explain recent measurement of interference in anti-dot based interferometers
at filling factor $\nu=2/3$.

\section*{ACKNOWLEDGMENTS}
We dedicate this article to the memory of Markus Buttiker. His valuable and insightful contributions to the study of edge states in the quantum Hall effect and his long lasting involvement in theoretical developments in the field make us hope that he would find the subject we present of interest.

We are grateful to Angela Kou and Charles Marcus for instructive discussions.
This work was supported by Microsoft's Station Q, the US-Israel Binational Science Foundation, the Minerva Foundation, the DFG grants RO 2247/7-1 and RO 2247/8-1 and the European Research Council under the European Community's Seventh Framework Program (FP7/2007-2013)/ERC Grant agreement No. 340210.



\end{document}